\documentclass[a4paper]{jpconf}

\usepackage{graphics}
\usepackage{graphicx}
\usepackage{bm}
\usepackage{amsmath}
\usepackage{amsfonts}
\usepackage{amssymb}
\usepackage{latexsym}
\usepackage{color}

\newcommand{\ket}[1]{| #1 \rangle}

\definecolor{darkred}{rgb}{.8,0,0}

\definecolor{darkblue}{rgb}{0,0,.8}

\newcommand{\be}{\begin{equation}}\newcommand{\ee}{\end{equation}}
\newcommand{\bea}{\begin{eqnarray}}\newcommand{\eea}{\end{eqnarray}}
\newcommand{\brr}{\begin{array}}\newcommand{\err}{\end{array}}
\newcommand{\bit}{\begin{itemize}}\newcommand{\eit}{\end{itemize}}
\newcommand{\ben}{\begin{enumerate}}\newcommand{\een}{\end{enumerate}}

\def\1{{_{1}}}\def\2{{_{2}}}

\def\be{\begin{equation}}
\def\ee{\end{equation}}

\def\bea{\begin{eqnarray}}
\def\eea{\end{eqnarray}}

\begin{document}

\title{Axion like particles, fifth force and neutron interferometry}

\author{Antonio Capolupo}
\address{Dipartimento di Fisica ``E.R. Caianiello'' Universit\`{a} di Salerno, and INFN --- Gruppo Collegato di Salerno, Via Giovanni Paolo II, 132, 84084 Fisciano (SA), Italy}

\author{Salvatore Marco Giampaolo}
\address{Institut Ruder Boskovic, Bijenicka cesta 54, 10000 Zagreb, Croatia}

\author{Aniello Quaranta}
\address{Dipartimento di Fisica ``E.R. Caianiello'' Universit\`{a} di Salerno, and INFN --- Gruppo Collegato di Salerno, Via Giovanni Paolo II, 132, 84084 Fisciano (SA), Italy}

\begin{abstract}

We report on recent result according to which the fermion-fermion interaction mediated by axions and axion--like particles can be revealed by means of  neutron interferometry. We assume that  the initial neutron beam is split in two  beams which are affected by differently oriented magnetic fields, in order that the phase difference depends only by the axion--induced interaction. In this way, the  phase difference is directly related to the presence of axions.

\end{abstract}

Phenomena, ranging from neutrino mixing \cite{Mixing1,Mixing2,Mixing3,Mixing4,Mixing5,Mixing6,Mixing7} to the dark matter and energy  \cite{dark1,dark2}, to the muon g-2 anomaly \cite{Muon1,Muon2}, together with the strong CP problem (i.e. the absence of CP symmetry violation in strong interaction, solved by Peccei and Quinn \cite{Quinn1977,Peccei1977} by introducing pseudo--scalar particles known as axions \cite{Quinn1977,Peccei1977,Weinberg1978,Wilczek1978,Raffelt2007}) show the   necessity of physics beyond the standard model of particles \cite{Ellis2009}.
Axions and axion--like particles (ALPs) \cite{DFSZ1, DFSZ2, PDG2020, KSVZ1, KSVZ2}, with masses  from the ultralight \cite{Kim2016,Demartino2017} $m \simeq 10^{-22} \mathrm{eV}$ to the heavy axions \cite{Rubakov1997} $m \simeq 1 \ \mathrm{TeV}$, could represent a possible dark matter component \cite{Marsch2016}. Many experiments searched for ALPs   \cite{PVLAS,OSQAR,OSQAR2,ALPS,CAST,ADMX,QUAX,Capolupo2015,Capolupo2019}, however, up to now no evidence for ALPs has been found.
ALPs interact with the electromagnetic field, moreover, they are expected to play the role of a mediating boson in the fermion--fermion interaction \cite{Moody1984,Daido2017}. Therefore, different  experiments were designed to probe such interactions  \cite{Bezerra2014,Klimchitskaya2015,Klimchitskaya2017,Capolupo2020}.

Another extremely thriving field of physics is the neutron interferometry \cite{Rauch2015}. It  has allowed  to verify many theoretical effects, such as  the Sagnac effect \cite{Werner1979}, the geometric phase \cite{Allman1997,Wagh1990,Wagh1998} and the wave--particle duality.

Here, we report on a new approach to  detect   ALPs, which suggests the use of a   neutron interferometer in which two sub--beams are subject to  external magnetic fields of equal strength but different direction, as a device to reveal fermion-fermion interaction mediated by axions. \cite{Capolupo-as2021}.
Indeed, we show that a detectable  neutron phase difference, depending only by the axion--induced interaction between neutrons, can be achieved by setting the magnetic fields in the arms of the interferometer, one in the direction of propagation of the relative sub-beam and the other one orthogonally to the propagation.
We fix the experimental parameters  in order  that the phase difference depends only on the axion--mediated interaction and the contributions given by the other interactions are removed. Then we show how a neutron interferometer is sensible to  the presence of ALPs  in a significant portion of parameter space.

The Lagrangian describing the neutron-neutron interaction due to axion exchange is   given by~\cite{Moody1984,Daido2017}
\begin{equation}\label{YukawaInteraction}
  \mathcal{L}_{INT} = -  \sum_{j=1,2}i g_{aj} \phi \bar{\psi}_{j}  \gamma_{5} \psi_{j}
\end{equation}
with  $\phi$  the axion field, $\psi_1,\psi_2$    fermion fields and $g_{aj}$   the (dimensionless) effective axion--fermion coupling constants, $g_{aj}=g_{aNN}$ in the case of axion-neutron coupling. Such constant is depending on the axion (or ALP) model and on the fermions interacting. We shall assume that the neutron velocities are non-relativistic and analyze this interaction potential within the context of ordinary quantum mechanics. In the non--relativistic limit $\mathcal{L}_{INT}$ yields a two--body potential for the neutrons \cite{Moody1984,Daido2017}. This axion--induced interaction is not the only one in play. However, the gravitational interaction is easily seen to be irrelevant, due to the smallness of the masses involved. In addition, we shall always assume a relative distance $r > 10^{-12} m$ among the neutrons, so that the short--range nuclear interactions can also be ignored. With this assumption the only other relevant interaction is the magnetic one between the neutron dipoles.
The neutron--neutron interaction Hamiltonian comprising the magnetic and the axion--mediated interaction, generalized to an arbitrary number of neutrons can  be written as \cite{Capolupo-as2021}
\begin{eqnarray}\label{Total_Hamiltonian3}
  H_{ij}  =  -\frac{\mathcal{A}}{r^3_{ij}} \bigg[ \left(3-\mathcal{B}e^{-mr_{ij}} K^{(a)}(r_{ij})\right) \sigma^{r_{ij}}_i \sigma^{r_{ij}}_j - \left(1-\mathcal{B}e^{-mr_{ij}}K^{(b)}(r_{ij})\right) \pmb{\sigma}_i \cdot \pmb{\sigma}_j \bigg] \ .
\end{eqnarray}
In eq. \eqref{Total_Hamiltonian3} the two contributions are signaled by the parameters  $\mathcal{A}=\frac{g^2 \alpha}{16 M^2}$, denoting the strength of the magnetic interaction ($g$ is the neutron g--factor, $M$ is the neutron mass and $\alpha$ is the fine-structure constant) and $\mathcal{B}=\frac{ g_{aNN}^2 }{ \pi \alpha g^2}$, representing
the relative weight of the axion interaction, which vanishes in absence of the ALP (whose mass is denoted $m$). The dimensionless (in natural units $c=1=\hbar$) functions $K(r)$ are defined as $K^{(a)}(r) = m^2 r^2 + 3m r + 3$, and $K^{(b)}(r) = mr + 1$. The vector $\pmb{r_{ij}} = \pmb{r_i} - \pmb{r_j}$ denotes the relative position of the neutrons $i$ and $j$, $r_{ij}=|\pmb{r_{ij}}|$ is their relative distance and the operators $\sigma^{r_{ij}}_l = \pmb{\sigma}_l \cdot \pmb{\hat{r}_{ij}}$ are defined by the projection of the pauli operators $\pmb{\sigma_l}$ of neutron $l$ on $\pmb{r_{ij}}$. The notation $\sigma^{r_{ij}}_i$ is used to remark that while these operators act only upon the space of the $i$-th particle, their form depends on the specific pair $i,j$ considered.  Eq.\eqref{Total_Hamiltonian3} can be recast in a more compact form by defining the functions $ C (r) = \frac{\mathcal{A}}{r^3}\left(1 - \mathcal{B} e^{-mr} K^{(b)}(r) \right)$, $D(r) = \frac{\mathcal{A}}{r^3}\left(3 - \mathcal{B} e^{-mr} K^{(a)}(r) \right)$ and the symmetric matrix $ K^{uv} (\pmb{r})= C (r) \delta^{uv} - D (r) R^{u} (\pmb{r}) R^{v} (\pmb{r})$ for $u,v = x,y,z$. The symbol $R^{u}(\pmb{r}) = \pmb{\hat{r}} \cdot \pmb{\hat{u}}$ denotes the projection of the vector $\pmb{r}$ on the $u$ axis. Thus $H_{ij} = \sum_{u,v} K^{uv}(\pmb{r}_{ij}) \sigma_{i}^{u} \sigma_{j}^{v}$. The two--neutron interaction can easily be generalized to an arbitrary number of neutrons. The total interaction Hamiltonian is simply the sum over all pairs $i,j$
$
  H = \frac{1}{2}\sum_{i,j} H_{ij}
$, where the factor $\frac{1}{2}$ accounts for double counting.

The study of the evolution of the $n$--neutron states   interacting via $H$ represents  a complicated many--body problem. However, we want to consider only   the evolution of the single neutron state and we are not interested to  correlations and collective effects. Therefore, by using a mean field approach, we describe the interaction of a neutron with all the other nucleons by means of an effective one--particle potential.
Let us consider a neutron at position $\pmb{r}_i$ and the Pauli operator $\pmb{\sigma}_i$, the instantaneous interaction hamiltonian due to the other neutrons is
$
  H_i = \sum_{u,v}  \sum_{j \neq i} K^{uv} (\pmb{r}_{ij}) \sigma_{j}^{u} \sigma^{v}_{i} \ ,
$
where the sum is relative to all the other neutrons $j$.
Since we are interested to   an effective local potential for the single neutron, then we substitute  the above equation   with its expectation value on the state of the other nucleons. Notice that the term of a spin interacting with a magnetic field
$
  H_{i} = -\mu\left(\pmb{B_i} (\pmb{r_i})\right) \cdot \pmb{\sigma} \ ,
$
is obtained by setting
\begin{equation}
 \mu \pmb{B_i} (\pmb{r_i}) =-\sum_u   \sum_{j} K^{uv} (\pmb{r_{ij}}) \langle \sigma_{j}^{u} \rangle \,.
\end{equation}

The evolution of the single neutron state is analyzed by  determining the effective magnetic field for a particular spatial spin configuration, and then by inserting the one particle operator into the Schroedinger equation.

For our purposes, we shall consider cold neutron beams with specific requirements. First of all, we deal with collimated neutron beams and a small beam width of the order of $10 \ \mathrm{\mu m}$. Experimentally, such kind of beams  can be produced in different ways as shown in  \cite{Ott2015}. Neglecting the angular spread, we can assume that the beam is distributed with cylindrical symmetry around the beam axis $\pmb{\hat{y}}$, and, considering it  sufficiently thin, we can represent  the beam as a monodimensional system.
We will also assume that, by using   a monochromator, one can select  only neutrons around a given value of the kinetic energy $K$. The beam intensity is expected to decay as the neutron beam propagates. However, as a first instance, we consider the beam intensity constant and neglect the losses due to the propagation.

\begin{figure}[t]\centering
\begin{picture}(300,180)(0,0)
\put(-20,20){\resizebox{13.0 cm}{!}
{\includegraphics[width=\linewidth]{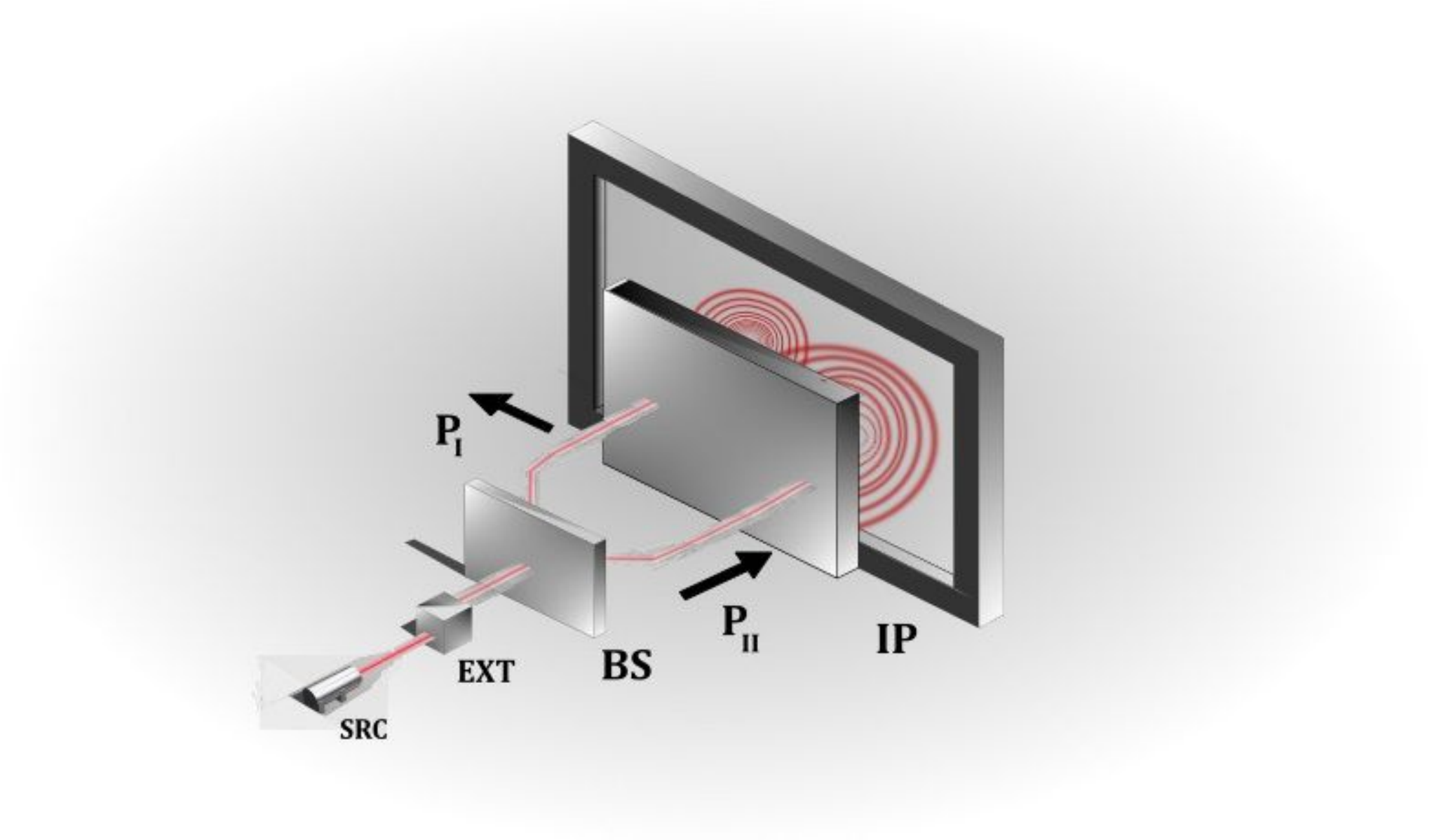}}}
\end{picture}
\vspace{-1cm}
\caption{
(color online) Diagram of the interferometric apparatus. The beam from the neutron source \textbf{SRC} is transmitted to a collimation device and a monochromator \textbf{EXT}. Then the   beam is   conveyed to a beam splitter \textbf{BS} which splits the beam into two subbeams with different average spin polarizations $\pmb{P}_I$ and $\pmb{P}_{II}$. In the end, the two subbeams interfere at the interference plane \textbf{IP}.
}
\label{IntDiagram}
\end{figure}

We now present, in Fig.(\ref{IntDiagram}), the scheme of the neutron interferometer eligible for  the detection of the axion--mediated interaction among fermions.
In such a scheme, a source, as a reactor, \textbf{SRC} generates cold or ultra--cold neutrons, which are directed to monochromator and a collimator \textbf{EXT}, making the beam (approximately) linear and monochromatic. Then   a beam splitter \textbf{BS}  separates the beam into two sub--beams $I$ and $II$ which enter  in a device  that selects two different spin polarizations $\pmb{P}_{I}$ and $\pmb{P}_{II}$. The sub--beams pass  through regions permeated by magnetic fields of the same strength but with distinct direction, i.e. $\pmb{B^0_J}=B_0 \pmb{P_J}$.

Making  as close as possible the  fractional intensities of the sub-beams $\chi_J$  with respect to the initial beam $I_0$:
$\chi_J = I_J /I_0$, $J = I, II$,  $\chi_I \simeq \chi_{II} $, one has that the average distances between successive neutrons in the two sub--beams
are similar $d_{I} \simeq d_{II}$. Indeed, $d_{J} = \frac{\bar{v}_J}{I_J}$, with $\bar{v}_J$   average neutron velocity   of propagation in the $J$ sub--beam. Moreover, considering constant $I_{J}$ and $\bar{v}_J$, the distances $d_{J}$ are constant, and the one particle Hamiltonian $H$ is time--independent. We assume that $d_J > 10^{-12} m$ in order that the effects of nuclear interactions are negligible.

 The interference is detected in a plane \textbf{IP} where the polarized sub--beams hit, after crossing two optical paths of the same length

We choose $\pmb{P}_{I}$ to be orthogonal to $\pmb{\hat{y}}_I$  and $\pmb{P}_{II}$ parallel to $\pmb{\hat{y}}_{II}$. Starting from the total interaction Hamiltonian, exploiting the assumption that all the distances between two subsequent neutrons in sub--beam $J$ are equal to $d_J$, with simple algebraic steps, one arrives at the effective magnetic fields \cite{Capolupo-as2021}. It is easy to show that, within these assumptions, each neutron in sub--beam $J$ is subject to the same effective magnetic field $\mu \pmb{B_J} = \mu B_J \pmb{P_J}$ given by
\begin{eqnarray}\label{SecondSeries}\nonumber
\! \! \! \! \!   \mu B_I \! &\!= \! & \! \!  - \frac{2\mathcal{A}\zeta(3)}{d^3_I} \! + \! \frac{2\mathcal{A} \mathcal{B} }{d^3_I}  \mathrm{Li}_{3} (e^{-md_I} \!) \! + \! \frac{2\mathcal{A} \mathcal{B} m}{d^2_I}  \mathrm{Li}_{2} (e^{-md_I}\!)\\
\mu B_{II} \! &\!=\!&\! \frac{4\mathcal{A}\zeta(3)}{d^3_{II}}- \frac{4\mathcal{A} \mathcal{B} }{d^3_{II}}  \mathrm{Li}_{3} (e^{-md_{II}}) -
  \frac{4\mathcal{A} \mathcal{B} m}{d^2_{II}}\mathrm{Li}_{2} (e^{-md_{II}})  \nonumber \\
&& + \frac{2 \mathcal{A} \mathcal{B} m^2}{d_{II}} \log(1 - e^{-md_{II}})
\end{eqnarray}
where $\zeta(s)$ stands for the Riemann zeta function while $\mathrm{Li}_{s}(z) = \sum_{n=1}^{\infty} \frac{z^n}{n^s}$ is the Polylogarythm function \cite{Gradshteyn}.
The Schroedinger equation ruling  the evolution of the single neutron state, in both sub--beams, is
\begin{equation}\label{Schrodinger}
  i \partial_t \psi_J = \left(- \frac{\nabla^2}{2M} + M \right) \psi_J - \pmb{\sigma} \cdot \left[\mu (\pmb{B_J} + \pmb{B^{0}_J} ) \right]\psi_J
\end{equation}
with $\psi_J$  the product of a spatial wave-function  and a spin function. We express the spinor for each sub--beam in the basis defined by the corresponding polarization, i.e.
$  \pmb{\sigma} \cdot \pmb{P_J} \ket{\uparrow_J} =  \ket{\uparrow_J}$ and $\pmb{\sigma} \cdot \pmb{P_J}\ket{\downarrow_J} = -\ket{\downarrow_J}.$
Moreover, we assume that the neutron, after passing the beam splitter,   is   in the up state for the corresponding sub--beam. We consider $t=0$ at this instant.
 Denoting with  $y$   the coordinate along the propagation axis, with $y=0$ and $t=0$ at the beginning of the optical path, one has
 $ \psi_J (t) = f(t) e^{i k y} \ket{\uparrow_J}
$
with $f(t)$   a function which can be assumed as  $f(t) = e^{- i \omega_J t}$. Then, from Eq.\eqref{Schrodinger} we have
$
  \omega_J = \frac{k^2}{2M} + M - \mu B_J - \mu B_0 \ ,
$
and the total phase   at time $t$ is given by
$
  \phi_J (t) = \arg \left(\langle \psi_J(0)| \psi_J (t) \rangle\right) = - \left(\frac{k^2}{2M} + M - \mu (B_J + \mu B_0) \right)t.
$
The phase difference between the two beams at time $t$ is
$
  \Delta \phi (t) = \phi_{II} (t) - \phi_{I}(t) =\mu \left( B_{II} -  B_{I} \right)t,
$
where $\mu   B_{I} $ and  $\mu   B_{II} $ are given in Eq. \eqref{SecondSeries}. By setting  $d_{I}=d_{II} = d$, $ \Delta \phi (t)$ can be written as $ \Delta \phi (t) = \left[G_{m} (d) + G_{a} (d)\right] t$,
where $ G_{m} (d)  = \frac{6\mathcal{A}}{d^3} \zeta (3)$ is due to the dipole--dipole interaction and
\begin{eqnarray}\label{ThirdSeries}
\nonumber
G_{a} (d) =  -\frac{6\mathcal{A} \mathcal{B} }{d^3}  \mathrm{Li}_{3} (e^{-md}) - \frac{6\mathcal{A} \mathcal{B} m}{d^2}\mathrm{Li}_{2} (e^{-md})
+ \frac{2 \mathcal{A} \mathcal{B} m^2}{d} \log (1-e^{-md})  \,, \nonumber
\end{eqnarray}
is due to the axion--mediated interaction.
Here the $n \rightarrow \infty$ limit is intended.

Neglecting a possible phase shift due to the beam splitter, the phase difference induced by the dipole-dipole interactions can be removed by setting the beam path in such a way that $ G_{m} (d)$  is an integer multiple of $2 \pi$.
  This is obtained for time intervals $T_k$, which  for any integer $k$  are
$
  T_k = \frac{2 k \pi}{G_{m} (d)}= \frac{k \pi d^3}{3 \mathcal{A} \zeta(3)} \ .
$
The phase difference, evaluated at $T_k$, is then
\begin{eqnarray}\label{Reduced Phase Difference 2}
  \Delta \phi (T_k)  &=&   \bigg\{\frac{k \pi \mathcal{B}}{3 \zeta(3)} \Big[ 2 m^2 d^2 \log(1-e^{-md})
  - 6 m d \mathrm{Li}_2(e^{-m d}) - 6 \mathrm{Li}_3(e^{-md}) \Big]\bigg\}_{\mathrm{mod} \,2 \pi},
\end{eqnarray}
and it is different from zero only in presence of ALPs.  Indeed, $\Delta \phi (T_k)= 0$   for $\mathcal{B}=0$.
  Eq.\eqref{Reduced Phase Difference 2} shows that $\Delta \phi (T_k)  $   is proportional to the parameter $\mathcal{B} \propto g_p^2$. For ALP  masses $m \in [10^{-6}-1]\mathrm{eV}$, and   distances $d \in [10^{-11}-10^{-6}]\mathrm{m}$, one has that $\Delta \phi (T_k)  $ depends only weakly on  $m$ and $d$, while it depends  strongly   on the coupling.
\begin{figure}[h]
\centering
\begin{picture}(300,180)(0,0)
\put(10,20){\resizebox{8.0 cm}{!}
{\includegraphics[width=0.94\linewidth]{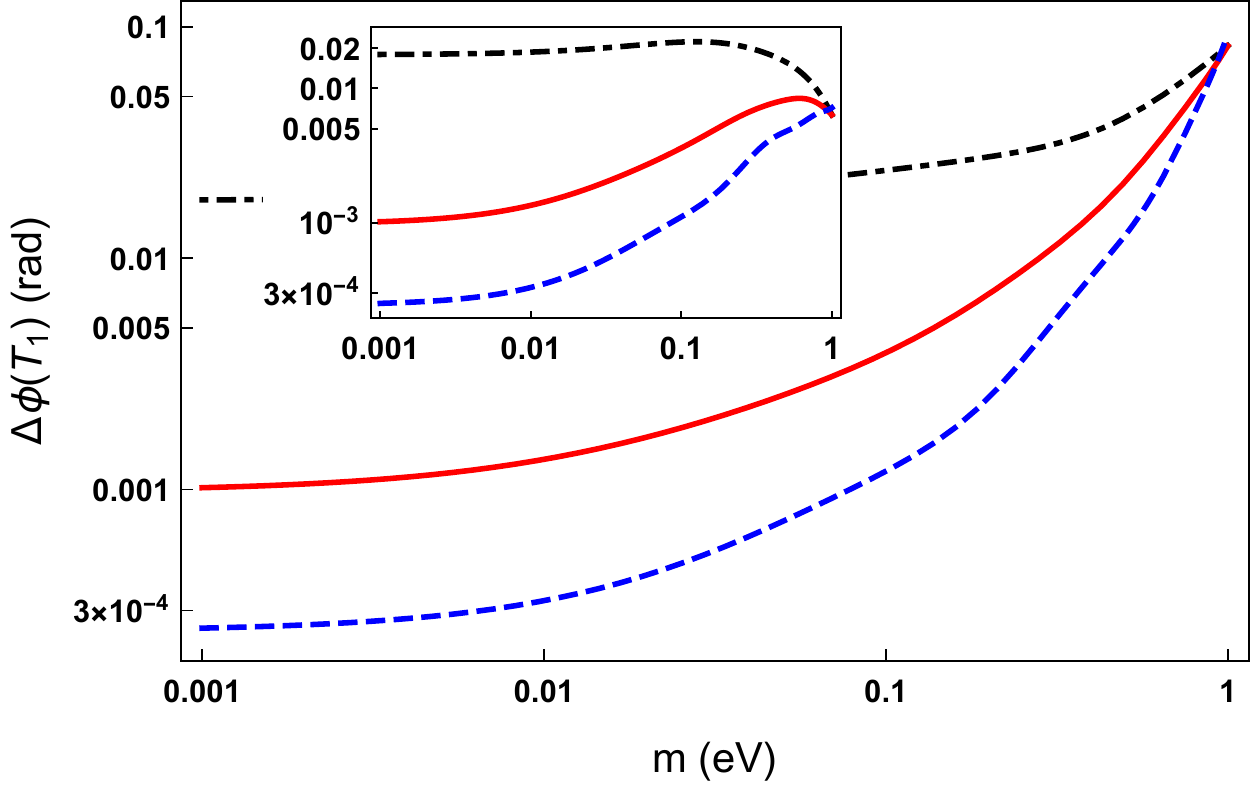}}}
\end{picture}\vspace{-1cm}
\caption{
(color online) Logarithmic scale plot of the phase difference $|\Delta \phi (T_{1})|$   modulo $2 \pi$ for several values of the coupling constant in the ALP mass range $[10^{-3},1] \ \mathrm{eV}$ for an inter--neutron distance $d=10^{-8} \mathrm{m}$ ($d=10^{-6} \mathrm{m}$ in the inset). In particular:   the black dot--dashed line is obtained by considering the threshold from effective Casimir pressure measurements ~\cite{Bezerra2014}, $g_p = g_{CP}$, and   using the following values $g_{CP} = 0.0327$ for $m = 10^{-3} \mathrm{eV}$, $g_{CP} = 0.0348$ for $m = 0.05 \mathrm{eV}$, $g_{CP} = 0.0674$ for $m = 1 \mathrm{eV}$. For the red solid line, we used  $g_p = g_{CF}$, where $g_{CF}$ is the threshold from measurements of the difference of Casimir forces ~\cite{Klimchitskaya2017} and the sample values are $g_{CF}=0.007$ for $m = 10^{-3} \mathrm{eV}$, $g_{CF} = 0.012$ for $m = 0.05 \mathrm{eV}$, $g_{CF} = 0.066$ for $m = 1 \mathrm{eV}$.  In the blue dashed line we assume $g_p = g_{IE}$, where $g_{IE}$ is the threshold from isoelectronic experiments~\cite{Klimchitskaya2015}, and sample values are $g_{IE} = 0.0036$ for $m = 10^{-3} \mathrm{eV}$, $g_{IE}=0.006$ for $m = 0.05 \mathrm{eV}$, $g_{IE}=0.07$ for $m = 1 \mathrm{eV}$.
}
\label{Moste}
\end{figure}
The phase difference, modulo $2 \pi$, is plotted in Fig.(2),  at the minimum recurrence time $T_{1}$, for several values of the coupling constant $g_p$ in the mass range $[10^{-3},1] \ \mathrm{eV}$. For axion masses $m < 0.1 \mathrm{eV}$, one has that  $\Delta \phi$   is almost independent on the distance   for $d=10^{-8} \mathrm{m}$ and $d=10^{-6} \mathrm{m}$.
The dependence on the distance is relevant only when the product $m d$ is quite high.
On the other hand, for $md \ll 1$, $\Delta \phi$ essentially depends only  on the coupling $g_p$.

\begin{figure}
\centering
\begin{picture}(300,180)(0,0)
\put(-80,20){\resizebox{16.0 cm}{!}{
\includegraphics[width=\linewidth]{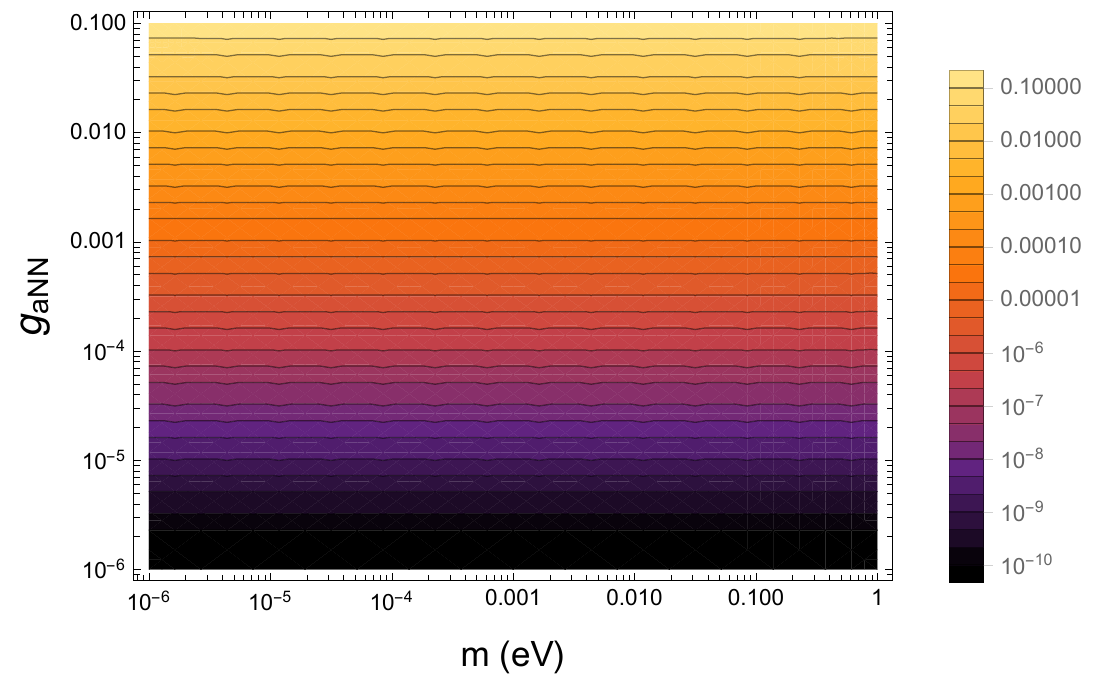}
\includegraphics[width=\linewidth]{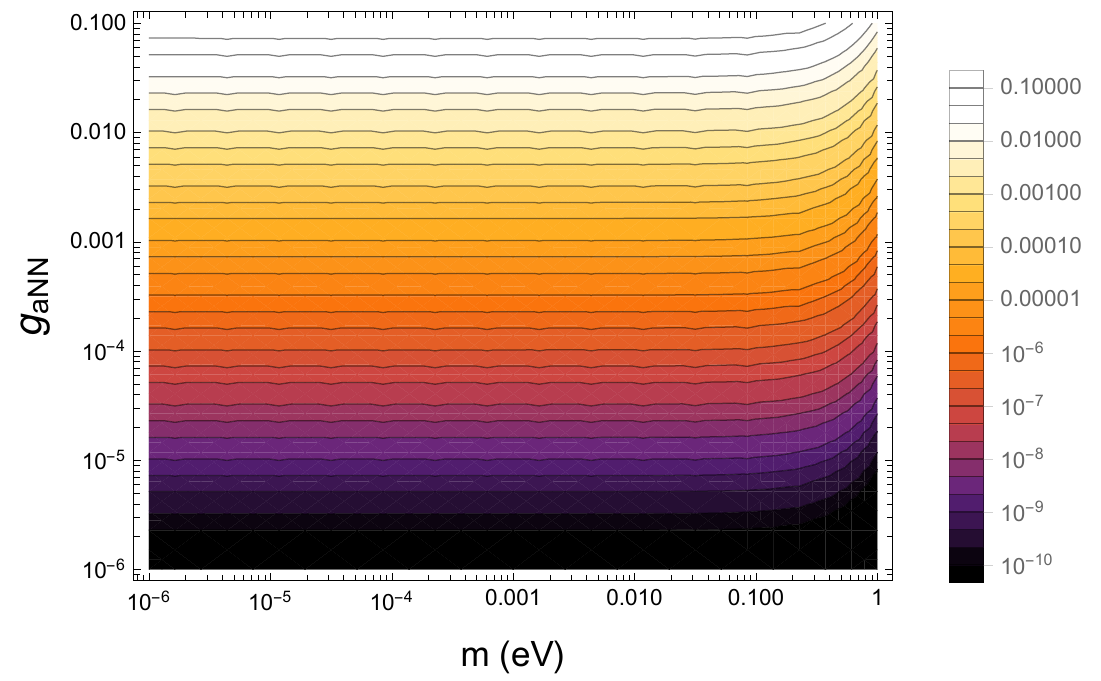}}}
\end{picture}\vspace{-1cm}
\caption{
(color online) Contour plots of  $|\Delta \phi (T_{1})|$ in the mass--coupling plane for $d=10^{-9} m$ (left panel) and $d=10^{-6} m$ (right panel).  The range of axion masses and coupling constant adopted is $(m,g_p) \in [10^{-6},1] \mathrm{eV} \times [10^{-6},10^{-1}]$.
}
\label{Contour}
\end{figure}
The plots of Fig.(\ref{Contour})  show   the $g_p^2$ dependence of $\Delta \phi (T_{1})$  and an extremely weak dependence on the ALP mass in the ranges analyzed.

Notice that in order to detect  the phase difference, one has to take into account the limitation coming from the need for a recurrence time $T$ required to isolate the axion contribution to $\Delta \phi  $, and the limitation due to the total neutron flux available, which is fundamental for a sufficient small inter-neutron distance.
The first limitation can be overcame by using  ultra cold neutrons  with velocities as small as $5 \ \mathrm{m/s}$ \cite{Steyerl1969}. These neutrons allow  to have minimal recurrence times of order of the second (less than the  neutron lifetime and the coherence time) and length of the interferometer of order of meters.
The second limitation can be passed by using neutron beams of intensities of the order $I \simeq (10^{8} - 10^{10}) n/s$ \cite{IAEA2014}.
Indeed, for neutron velocity   of the order of $v \simeq 1 \ \mathrm{m/s}$, a distance
   $d \simeq 10^{-8} \mathrm{m}$ is obtained in correspondence with an intensity $I \simeq 10^{8} \mathrm{n/s}$.

In conclusion, we have  shown that, in the time evolution of neutrons, the axion mediated fermion-fermion interaction
generates a  contribution to  the total phase. Such a contribution can be in principle detectable, by using ultra cold neutrons in an interferometer
in which the  two  beams  are affected by differently oriented magnetic fields, in order that the phase difference depends only by the axion--induced interaction. Significant  phase difference is obtained for a wide range of ALP parameters.

\section*{Acknowledgements}

A.C. and A.Q thank  partial financial support from MIUR and INFN. A.C. also thanks the COST Action CA1511 Cosmology and Astrophysics Network for Theoretical Advances and Training Actions (CANTATA).
SMG acknowledge support from the European Regional Development Fund for the Competitiveness and Cohesion Operational Programme (KK.01.1.1.06--RBI TWIN SIN) and from the Croatian Science Fund Project No. IP-2016--6--3347 and  IP-2019--4--3321.
SMG also acknowledge the QuantiXLie Center of Excellence, a project co--financed by the Croatian Government and European Union through the European Regional Development Fund--the Competitiveness and Cohesion Operational Programme (Grant KK.01.1.1.01.0004).

\end{document}